\documentstyle[12pt]{article}
\def\beq{\begin{equation}}
\def\eeq{\end{equation}}
\def\bea{\begin{eqnarray}}
\def\eea{\end{eqnarray}}

\def\gappeq{\mathrel{\rlap {\raise.5ex\hbox{$>$}}
{\lower.5ex\hbox{$\sim$}}}}
\def\lappeq{\mathrel{\rlap{\raise.5ex\hbox{$<$}}
{\lower.5ex\hbox{$\sim$}}}}

\def\all{a_\mu^{\rm EW} ({\rm 2\, loop})_{ LL}}
\def\amu{a_\mu}
\def\aew{a_\mu^{\rm EW}}
\def\sw{s_W^2}

\begin{document}
\begin{titlepage}
\begin{center}
March 1998  \hfill    CERN-TH/97-86\\
               \hfill    hep-ph/9803384
\vskip .2in
{\large \bf 
QED Logarithms in the Electroweak Corrections to the Muon Anomalous
Magnetic Moment}

\vskip .3in

\vskip .3in
G. Degrassi\footnote{Permanent address: Dipartimento di Fisica, Universit{\`a}
                  di Padova, Padova, Italy.} and
G.F. Giudice\footnote{On leave
of absence from INFN, Sez. di Padova, Italy.}

\vskip .2in

{\em Theory Division, CERN\\
     CH-1211 Geneva 23, Switzerland}

\end{center}
\vskip .2in
\begin{abstract}
\medskip

We employ an effective Lagrangian approach to derive the 
leading-logarithm 
two-loop electroweak contributions to the muon anomalous magnetic 
moment, $\amu$. 
We show that these corrections can be obtained using known results on
the anomalous dimensions of
composite operators.
We confirm the result
of Czarnecki {\it et al.} for the bosonic part and present the complete
$\sin^2 \theta_W$ dependence of the fermionic contribution.
The approach is then used to compute the leading-logarithm 
three-loop  electroweak contribution to $\amu$. 
Finally we derive, in a fairly model-independent way, 
the QED improvement of new-physics contributions to
$\amu$ and to the electric dipole moment (EDM)
of the electron. We find that the
QED corrections reduce the effect of new physics at the electroweak scale
by 6\% (for $\amu$) and by 11\% (for the electron EDM).

\end{abstract}

\vskip 3cm

\end{titlepage}

\newpage
\baselineskip20pt

\section{Introduction}

Historically, the measurement of the anomalous magnetic moment of the 
muon~\cite{pdg}
\beq
a_\mu\equiv \frac{g_\mu -2}{2}\equiv \frac{\mu_\mu}{(e\hbar /2m_\mu )}
-1 = (11~659~230 \pm 84 )\times 10^{-10}
\label{exp}
\eeq
has provided the most convincing test of the validity 
of QED~\cite{mar}. 
As the precision in the experimental measurement and in the theoretical
determination are progressively improving, the muon anomalous magnetic
moment $\amu$ may soon offer a test of the Standard Model (SM) electroweak
theory and of some of its possible extensions.

From the experimental point of view, the E821 experiment at Brookhaven
National Laboratory is expected to improve the accuracy in the $\amu$
measurement to the level of
$4\times 10^{-10}$, and possibly to 1--2$\times 10^{-10}$ if large
statistics is accumulated \cite{bun}. Let us consider now
the present status of the
theoretical prediction for $\amu$. It is convenient to separate the
total result
into several different parts.
The pure QED contribution, which is known to order $\alpha^5$~\cite{mar}, is
\beq
a_\mu^{\rm QED}=(116~584~706\pm 2)\times 10^{-11}~.
\eeq
The part affected by strong-interaction 
contributions, which contains the largest
source of uncertainty, comes from the hadron vacuum polarization 
and the hadronic light-by-light amplitude. 
A recent
analysis relating, by means of dispertion relations, the 
hadronic vacuum polarization to data from
$e^+e^-$ annihilation and $\tau$ decays gives~\cite{dav}
\beq
a_\mu^{\rm had}({\rm vac~pol})=(6951\pm 75)\times 10^{-11}~.
\eeq
The error is dominated by the experimental uncertainty and will be 
significantly improved by future measurements at BEPC in
Beijing, at DA$\Phi$NE in Frascati, and at VEPP-2M in Novosibirsk.
The effect of 
higher-order hadronic contributions have also been evaluated~\cite{Krau}
\beq
a_\mu^{\rm had}({\rm h.o. \,vac~pol})=(-101 \pm 6)\times 10^{-11}~.
\eeq
The most recent theoretical estimate of the
hadronic light-by-light contribution gives~\cite{hay}
\beq
a_\mu^{\rm had}(\gamma
\times \gamma ) =(-79\pm 15)\times 10^{-11}~.
\eeq
Finally, the SM electroweak contribution at one loop is
\cite{tutti}
\beq
a_\mu^{\rm EW} ({\rm 1\,loop}) =\frac{5\,G_\mu \, m_\mu^2}{24\sqrt{2}\pi^2}
\left[ 1+\frac{1}{5}(1-4\sw )^2\right] = 195\times 10^{-11}~,
\eeq
where $G_\mu$ is the Fermi constant and $\sw \equiv \sin^2\theta_W =
1-M_W^2/M_Z^2$. As first noticed in ref.~\cite{russi}, 
two-loop electroweak corrections  are quite substantial, because of
large contributions $O(G_\mu \, m_\mu^2\, \alpha / \pi \ln\, (M/m_f))$.
Here $M$ represents the $W$ or $Z$ mass and $m_f$ 
indicates a light fermion mass. 
Combining together the contribution containing 
closed fermionic loops~\cite{marci1,deraf} and the one from
the other relevant 
two-loop diagrams (usually indicated as bosonic part),
Czarnecki {\it et al.} obtain~\cite{marci2}
\beq
a_\mu^{\rm EW} ({\rm 2\, loop}) = (-44\pm 4)\times 10^{-11}~,
\eeq
for an Higgs mass $M_H= 250$ GeV, 
where the error  is associated to the uncertainties in $M_H$, the
hadronic contributions, and higher-loop effects.

In this paper, we show how the leading 
$O(G_\mu \, m_\mu^2\,\alpha / \pi \ln\, (M/m_f))$
two-loop electroweak corrections to the anomalous magnetic moment, $\all$,
can be easily obtained with the help of effective theories and
Wilson renormalization group. 
The relevant anomalous dimensions of composite operators can be extracted
from known results 
of the QCD corrections to flavour-violating bottom-quark transitions
mediated by the magnetic dipole operator~\cite{bur}. 
In this way, we derive 
$\all$ without directly computing any Feynman diagram and confirm the
results of refs.~\cite{marci1,deraf,marci2}.
Moreover, since the renormalization-group technique actually includes
all leading QED logarithms, we are able to give an analytic expression
for the
leading three-loop contribution to $\aew$. This contribution
turns out to be very small, but its knowledge allows to eliminate the
uncertainty on $\aew$ from higher-order effects. Including the different
terms, we obtain
\beq
\aew =(153\pm 3)\times 10^{-11}~,
\label{aweak}
\eeq
where the central value corresponds to  $M_H=150$ GeV.
This leads to a theoretical prediction for the muon anomalous
magnetic moment
\beq
\amu =(116~591~630\pm 77)\times 10^{-11}~.
\eeq

We also show that this method is well suited to compute higher-order 
corrections to new-physics contributions to magnetic and electric dipole
moments. 
The only necessary basic assumption is that the new physics gives a
one-loop contribution to the dipole moments, but does not significantly
affect at tree level composite four-fermion interactions. Under
this assumption, the leading logarithmic two-loop contribution is
determined by infrared effects and it can be computed from the
renormalization-group evolution of the effective theory below the
weak scale. We find that the QED improvement reduces the one-loop 
new-physics effect by 6\% in the case of $\amu$ and 11\% in the case
of the electric dipole moment (EDM) 
of the electron, assuming that the new physics
lies around the weak scale.

\section{Two-Loop Calculation of $\aew$}


The leading two-loop contributions to $\aew$ come from large QED
logarithms. These terms correspond to ultraviolet 
divergences in the effective theory obtained
by integrating out the heavy modes, and they can be derived in 
terms of the anomalous
dimensions of the dipole and current-current operators. 

We start by 
defining an effective theory valid below the electroweak scale, in which
the $W$ and $Z$ bosons and the top quark have been integrated out. The
effect from heavy particles is reflected in higher-dimensional operators.
Here we are interested in the operator corresponding to $\amu$ 
\beq
H_\mu=-\sqrt{\frac{\alpha}{(4\pi)^3}}~m_\mu ~{\bar \mu}\,\sigma^{\nu \rho} \,
\mu ~ F_{\nu \rho}~,
\label{hmu}
\eeq
and to other possible dimension-six operators that mix under QED 
renormalization with $H_\mu$. Since QED is parity conserving and $H_\mu$
is parity even, it is
easy to realize that we need to consider only
the following parity-even four-fermion operators
\beq
V_\mu =\frac{1}{2}~ \bar \mu \gamma^\nu \mu ~ \bar \mu \gamma_\nu \mu ~,~~~~~
A_\mu =\frac{1}{2}~ \bar \mu \gamma^\nu \gamma_5 \mu ~ \bar \mu \gamma_\nu
\gamma_5 \mu ~,
\eeq
\beq
V_{\mu f} = \bar \mu \gamma^\nu \mu ~\bar f \gamma_\nu f ~,~~~~~
A_{\mu f} = \bar \mu \gamma^\nu \gamma_5 \mu ~\bar f \gamma_\nu
\gamma_5 f ~.
\label{opp}
\eeq
In eq.~(\ref{opp}) $f$ indicates a generic fermion different from $\mu$ and
the factor  $1/2$ in the definition of the operators $V_\mu$ and $A_\mu$ 
compensates the symmetry factor for two identical currents in the Feynman
rule.
The relevant part of the effective Lagrangian is
\beq
{\cal L}_{\rm eff} =-2\sqrt{2}G_\mu \,\sum_i C_i {\cal O}_i ~,
\eeq
where the operators ${\cal O}_i$ are given in eqs.~(\ref{hmu})--(\ref{opp})
(with $f= \{ b,\tau ,c,s,d,u,e \} $)
and the Wilson coefficients $C_i$ at the electroweak scale 
 $\mu =M$ are
\beq
C_{H_\mu}(M)= \frac{2\sqrt{2}\,\pi^2}{G_\mu\,m_\mu^2}a_\mu^{\rm EW}
        ({\rm 1\,loop})
=\frac{5}{12}\left[ 1+\frac{1}{5}(1-4\sw )^2\right] ~,
\label{chmu}
\eeq
\beq
C_{V_\mu}(M) = \frac{1}{8}(1-4\sw )^2 ~,~~~~~
C_{A_\mu}(M) = \frac{1}{8}~,
\label{cmu}
\eeq
\beq
C_{V_{\mu f}}(M) = -\frac{1}{4}(1-4\sw )(T_f-2Q_f\sw )~,~~~~~
C_{A_{\mu f}}(M) = -\frac{1}{4}T_f~.
\label{cmuf}
\eeq
In eq.~(\ref{cmuf}) $T_f$ and $Q_f$ are the third isospin component and
electric charge of the fermion $f$, respectively.
The coefficient $C_{H_\mu}$ is obtained by one-loop integration, while
the coefficients in eqs.~(\ref{cmu})--(\ref{cmuf}) correspond to
tree-level $Z$ exchange. The $W$ boson can only generate operators with
$f=\nu$ which are irrelevant for our analysis, because they cannot mix
under QED with $H_\mu$ as neutrinos carry no electric charge.

We are interested in the Wilson coefficient of the operator $H_\mu$
at the scale $\mu= m_\mu$. The leading-log evolution of the Wilson
coefficients from the scale $M$ to a generic scale $\mu$
is given in terms of
the one-loop anomalous dimension matrix $\gamma $ and the one-loop 
beta function. In our case,  
\bea
C_i(\mu )&=&\sum_j\left[ \exp 
\int_{e(M)}^{e(\mu)} de~ \frac{\gamma^T(e)}{\beta(e)}
\right]_{ij} C_j(M)\nonumber \\
&=&\sum_j\left\{ V\left[ \frac{\alpha (M)}{\alpha(\mu)}\right]^{
\hat{ \frac{\gamma}{2b}}} V^{-1}\right\}_{ij} C_j(M)
~,
\label{rge}
\eea
where $e=\sqrt{4\pi \alpha}$ is the QED gauge coupling, and the 
rotation matrix $V$ is defined such that $\widehat \gamma =V^{-1} \gamma^T V$ is
diagonal. The beta function is  $\beta(e) = - b\,e^3/(16\pi^2) $,
with the coefficient $b$ given by
\beq
b =  -\frac{4}{3} \sum_f N_f Q_f^2~.
\label{beta}
\eeq
In eq.~(\ref{beta}) the sum is extended over all fermions $f$ with mass less 
than the
scale $\mu$, electric charge $Q_f$ and  multiplicity $N_f$
($N_f=3$ for quarks and $N_f=1$ for leptons).
Expanding eq.~(\ref{rge}) in powers of $\alpha$, we obtain
\beq
C_{H_\mu} (\mu)= C_{H_\mu} (M) -\sum_i \gamma({\cal O}_i, H_\mu)
\frac{\alpha(\mu)}{4\pi} \ln \frac{M}{\mu}C_{{\cal O}_i}(M)~,
\label{cfin}
\eeq
where $\gamma({\cal O}_i, H_\mu)\equiv \gamma_{{\cal O}_i H_\mu}$.
Aside from an overall factor $G_\mu\,m_\mu^2/(2\sqrt{2}\pi^2)$, 
the first term in the r.h.s. of
eq.~(\ref{cfin}) gives $a_\mu^{\rm EW} ({\rm 1\,loop})$, while the
second term gives  $\all$.

The last ingredient necessary to complete the analysis is the computation
of the elements $\gamma({\cal O}_i, H_\mu)$ of the anomalous dimension matrix.
This requires a calculation of the divergent parts of loop diagrams 
generating $H_\mu$, in which a single operator ${\cal O}_i$ is inserted and
a single photon is exchanged. Actually, this calculation is completely
analogous to the one of the QCD
anomalous dimension matrix for the $\Delta B=1$
effective Lagrangian, relevant for flavour-violating bottom-quark transitions.
For such a processes a  complete list of the divergent contributions from the 
various diagrams can be found in ref.~\cite{ciuco}. All we need to do is the 
proper translation from quarks to muons, and
from gluons to photons.  In this way we obtain
\beq
\gamma(H_\mu , H_\mu )=16
\label{ano1}
\eeq
\beq
\gamma(V_\mu , H_\mu )=\frac{40}{3}
\eeq
\beq
\gamma(A_\mu , H_\mu )=\frac{808}{9}
\eeq
\beq
\gamma(V_{\mu f} , H_\mu )=\frac{32}{9}\,Q_f^2N_f
\eeq
\beq
\gamma(A_{\mu f} , H_\mu )=48\,Q_f^2N_f~.
\label{ano2}
\eeq
These expressions are valid both in dimensional regularization with
the 't Hooft-Veltman prescription for $\gamma_5$~\cite{velt} and in 
dimensional reduction~\cite{dred}. In both schemes there are no 
finite operator renormalizations from non-vanishing matrix elements.

Replacing in eq.~(\ref{cfin}) the matching conditions of 
eqs.~(\ref{chmu})--(\ref{cmuf})
and the anomalous-dimension elements of eqs.~(\ref{ano1})--(\ref{ano2}),
and choosing $M = M_Z$ as high-energy scale, we obtain 
\bea
a_\mu^{\rm EW} ({\rm 2\,loop})_{LL}
& =&\frac{5 \,G_\mu\, m_\mu^2}{24\sqrt{2}\pi^2}
\frac{\alpha(m_\mu)}{\pi}\left\{ -\frac{43}{3} \left[ 1+\frac{31}{215}
(1-4\sw )^2\right] \ln \frac{M_Z}{m_\mu}\right. \nonumber \\
&+&\left. \frac{36}{5}\sum_{f\in F}N_f Q_f^2\left[ T_f+\frac{2}{27}
\left( T_f-2\, Q_f\,\sw \right)(1-4\sw )\right] \ln 
\frac{M_Z}{m_f}\right\}~,
\label{twol}
\eea
where the sum is extended over the fermions with a mass threshold between
$M_Z$ and $m_\mu$, $F=\{ b,\tau ,c,s,d,u\} $. Eq.~(\ref{twol}) 
confirms the results of
refs.~\cite{marci1,deraf,marci2}, but it disagrees with the one presented in 
ref.~\cite{russi}. It also shows the complete $\sw$ dependence,
extending the results of refs.~\cite{marci1,deraf}, in which the
fermionic contribution has been computed in the limit $\sw =1/4$.

The contribution from light quarks is not appropriately described by
eq.~(\ref{twol}), since perturbation theory is not justified. Nevertheless,
here we will parametrize the effect by taking in eq.~(\ref{twol})
$m_Q\equiv m_u =m_d=m_s=0.3$ GeV, and varying $m_Q$ between 0 and 1 GeV.
In this way, our result is consistent with the estimate of the 
light-quark contribution given in ref.~\cite{deraf}, based on  a chiral
effective Lagrangian. Taking
$m_c=1.5$ GeV and
$m_b=4.5$ GeV,
we find $\all =-(37\pm 1) 
\times 10^{-11}$. This corresponds
to $\sim 19\%$ of the one-loop contribution. We recall that the result in 
eq.~(\ref{twol})
includes only the logarithmic contribution. The terms not enhanced by
large logarithms have been computed in ref.~\cite{marci1,marci2} and 
amount to
$a_\mu^{\rm EW} ({\rm 2\,loop})_{NL} =-(6\pm 2)\times 10^{-11}$, where the
central value corresponds to a Higgs mass $M_H=150$ GeV and the error to a
variation of $M_H$ in the range between 100 and 1000 GeV. 

\section{Three-Loop Calculation of $\aew$}

In the previous section, we have computed $\all$
by expanding eq.~(\ref{rge}) at the first order in $\alpha$. However 
eq.~(\ref{rge}) also contains the information of higher-order terms,
since it resums all leading logarithms. We can therefore easily
evaluate the magnitude of  
higher-order corrections. For this purpose it is
 sufficient to expand eq.~(\ref{rge}) to the next order in $\alpha$, because
$(\alpha /\pi)\ln (M_Z/m_\mu)$ is much smaller than one. 
 The corresponding result will give us  the leading
logarithmic part of $a_\mu^{\rm EW} ({\rm 3\,loop})$.

Retaining ${\cal O}(\alpha^2)$ terms, the expansion of eq.~(\ref{rge}) yields
\beq
C_i (\mu)= \sum_j\left\{ \delta_{ij} -\gamma_{ji}
\frac{\alpha(\mu)}{4\pi} \ln \frac{M}{\mu}
+\left[ b\gamma_{ji} +\frac{1}{2} (\gamma \gamma )_{ji} \right]
\left[ \frac{\alpha(\mu)}{4\pi} \ln \frac{M}{\mu} \right]^2\right\}
C_j(M)~.
\label{trel}
\eeq
Because of the presence of the $(\gamma \gamma )$ factor, we now need 
information on the complete structure of the anomalous dimension matrix, and
not only on the elements given in eqs.~(\ref{ano1})--(\ref{ano2}). We start
by defining the basis for the required parity-even operators. 
Besides the operators defined in eqs.~(\ref{hmu})--(\ref{opp}), we
need the following four-fermion 
operators
\beq
V_f =\frac{1}{2} ~\bar f \gamma^\nu f ~\bar f \gamma_\nu f~,~~~~~
A_f =\frac{1}{2} ~\bar f\gamma^\nu \gamma_5 f~\bar f\gamma_\nu
\gamma_5 f~,
\label{opg}
\eeq
\beq
V_{f f'} = \bar f\gamma^\nu f~\bar f' \gamma_\nu f' ~,~~~~~
A_{f f'} = \bar f\gamma^\nu \gamma_5 f~\bar f' \gamma_\nu
\gamma_5 f' ~~~~{\rm with}~f\ne f'~,
\label{opf}
\eeq
\beq
{\widetilde V}_{q q'} = \bar q\gamma^\nu q'~\bar q' \gamma_\nu q ~,~~~~~
{\widetilde A}_{q q'} = \bar q\gamma^\nu \gamma_5 q'~\bar q' \gamma_\nu
\gamma_5 q ~~~~{\rm with}~q \ne q'~.
\label{opw}
\eeq
In eqs.~(\ref{opg})-(\ref{opw}) $f,\, f'$ $(q,\, q')$
represent a generic fermion (quark), 
and all quark operators are defined with the colour indices  saturated
so that each current is an $SU(3)_C$ singlet. The operators
${\widetilde V}_{q q'}$ and ${\widetilde A}_{q q'}$ cannot be written in terms
of $V_{f f'}$ and $A_{f f'}$ with a Fierz rearrangement, because of the
different colour-index saturation.

The matching conditions for the Wilson coefficients at the scale $\mu=M$ 
are
\beq
C_{V_{ f}}(M) = \frac{1}{2}(T_f-2\,Q_f\sw )^2~,~~~~~
C_{A_{ f}}(M) = \frac{1}{2}T_f^2~,
\label{cnew}
\eeq
\beq
C_{V_{f f'}}(M) = \frac{1}{2}(T_f-2\,Q_f\sw )(T_{f'}-2\,Q_{f'}\sw )~,~~~~~
C_{A_{f f'}}(M) = \frac{1}{2}T_fT_{f'}~,
\eeq
\beq
C_{{\widetilde V}_{q q'}}(M) = 
C_{{\widetilde A}_{q q'}}(M) =\frac{1}{4}\Delta_{qq'}~,
\label{cneww}
\eeq
where $\Delta_{qq'}=1$ if $q$ and $q'$ belong to the same isospin doublet,
and $\Delta_{qq'}=0$ otherwise. The coefficients for the operators of type
$V$, $A$ ($\widetilde V$, $\widetilde A$) are determined by tree-level $Z$ 
($W$) 
exchange and we neglect Cabibbo-Kobayashi-Maskawa angles.

The non-vanishing elements of the anomalous-dimension matrix 
for the operators in eqs.~(\ref{opg})-(\ref{opw})
are
\beq
\gamma ( V_f , V_f ) = \frac{16}{3} Q_f^2 (1+2N_f)
\label{ano3}
\eeq
\beq
\gamma ( V_f , A_f ) = 12 \,Q_f^2 
\eeq
\beq
\gamma ( A_f , V_f ) = \frac{52}{3} Q_f^2 
\eeq
\beq
\gamma ( V_{ff'} , V_{ff'} ) = \frac{16}{3} (Q_f^2 N_f + Q_{f'}^2 N_{f'})
\eeq
\beq
\gamma ( V_{ff'} , A_{ff'} ) = \gamma ( A_{ff'} , V_{ff'} ) =
12 \,Q_f Q_{f'}
\eeq
\beq
\gamma ( V_{f} , V_{ff'} ) = \frac{8}{3} Q_fQ_{f'}(1+2N_{f})
\eeq
\beq
\gamma ( A_{f} , V_{ff'} ) = \frac{8}{3} Q_fQ_{f'}
\eeq
\beq
\gamma ( V_{ff'} , V_{f} ) = \frac{32}{3} Q_fQ_{f'}N_{f'}
\eeq
\beq
\gamma ( V_{ff'} , V_{ff''} ) = \frac{16}{3} Q_{f'}Q_{f''}N_{f'}
\eeq
\beq
\gamma ( {\widetilde V}_{qq'} , {\widetilde V}_{qq'} ) = 
\gamma ( {\widetilde A}_{qq'} , {\widetilde V}_{qq'} ) = 
 12 \,Q_qQ_{q'}
\eeq
\beq
\gamma ( {\widetilde V}_{qq'} , V_{q} ) = 
\gamma ( {\widetilde A}_{qq'} , V_{q} ) = 
\frac{16}{3} Q_qQ_{q'}
\eeq
\beq
\gamma ( {\widetilde V}_{qq'} , V_{qq'} ) = 
\gamma ( {\widetilde A}_{qq'} , V_{qq'} ) = 
\frac{8}{3} (Q_q^2+Q_{q'}^2)
\eeq
\beq
\gamma ( {\widetilde V}_{qq'} , V_{qf''} ) = 
\gamma ( {\widetilde A}_{qq'} , V_{qf''} ) = 
\frac{8}{3} Q_{q'}Q_{f''}~,
\label{ano4}
\eeq
with $f'\ne f$, $q' \ne q$ and $f''\ne f,f'$.

Inserting in eq.~(\ref{trel})
the Wilson coefficients at the scale $\mu =M_Z$ (see
eqs.~(\ref{chmu})--(\ref{cmuf}) and eqs.~(\ref{cnew})--(\ref{cneww})),
and the relevant elements of the anomalous-dimension matrix 
(see eqs.~(\ref{ano1})--(\ref{ano2}) and eqs.~(\ref{ano3})--(\ref{ano4})),
we obtain
\beq
a_\mu^{\rm EW} ({\rm 3\,loop})_{LL}= \frac{5\,G_\mu\, m_\mu^2}{24\sqrt{2}\pi^2}
\left[ \frac{\alpha(m_\mu)}{\pi}\right]^2 \left( A+B \right) ~.
\eeq
Here $A$ and $B$ come respectively from the $(\gamma \gamma )/2$ and the
$b \gamma$ term in eq.~(\ref{trel}) and, in the approximation
$\sw = 1/4$ and $m_s = m_u = m_d = m_Q$, are given by
\bea
A &=&  \frac{2827}{90} \ln^2 \frac{M_Z}{m_\mu}  -
          \frac{298}{45} \ln^2 \frac{M_Z}{m_\tau} - 
          \frac{7826}{3645} \ln^2 \frac{M_Z}{m_b} +
           \frac{7040}{729}  \ln^2 \frac{M_Z}{m_c} + 
          \frac{2108}{405} \ln^2 \frac{M_Z}{m_Q}          \nonumber \\ &&
  + \frac{24}5  \ln \frac{M_Z}{m_b} \ln \frac{M_Z}{m_\mu}  
  + \frac{72}5  \ln \frac{M_Z}{m_\tau} \ln \frac{M_Z}{m_\mu} 
  - \frac{96}5  \ln \frac{M_Z}{m_c} \ln \frac{M_Z}{m_\mu}  
  - \frac{48}5  \ln \frac{M_Z}{m_Q} \ln \frac{M_Z}{m_\mu}   \nonumber \\ &&
  - \frac{128}{1215} \ln \frac{M_Z}{m_b} \ln \frac{M_Z}{m_c} 
\eea
\bea
B &=& - \frac{179}{45} \left( \frac13 \ln^2 \frac{M_Z}{m_b} +
          \ln^2 \frac{M_Z}{m_\tau} + \frac43 \ln^2 \frac{M_Z}{m_c} +
          2 \ln^2 \frac{M_Z}{m_Q} + 2\ln^2 \frac{M_Z}{m_\mu} \right) 
          \nonumber \\ &&
          + \frac25 \left(  \ln^2 \frac{m_b}{m_\tau} + 
          \frac43 \ln^2 \frac{m_b}{m_c} + 2 \ln^2 \frac{m_b}{m_Q} +
          2\ln^2 \frac{m_b}{m_\mu} \right) 
          - \frac85 \left(  2 \ln^2 \frac{m_c}{m_Q} +
          2\ln^2 \frac{m_c}{m_\mu} \right) \nonumber \\ && 
           +\frac65 \left( \frac43 \ln^2 \frac{m_\tau}{m_c} + 
           2 \ln^2 \frac{m_\tau}{m_Q} + 2\ln^2 \frac{m_\tau}{m_\mu} \right) -
          \frac85 \ln^2 \frac{m_Q}{m_\mu}
\eea

Using the same values for the quark masses as
in Sect.~2, we find 
\beq
\frac{a_\mu^{\rm EW} ({\rm 3\,loop})_{LL}}{\all} \simeq 
-0.8 \, \frac{\alpha}{\pi}\, \ln \frac{M_Z}{m_\mu} ~,
\eeq
which corresponds to a $1\%$ reduction of $a_\mu^{\rm EW} ({\rm 2\, loop})$
and gives $a_\mu^{\rm EW} ({\rm 3\,loop})_{LL} = 0.5 \times
10^{-11}$. Including all the different contributions, we obtain 
\beq
\aew =(153\pm 3)\times 10^{-11}~.
\eeq

\section{New Physics Effects in $\aew$ and in the Electron EDM}

The effective-Lagrangian method is well suited to discuss effects from
new physics. Indeed, in the presence of new interactions, characterized by
a mass scale $\Lambda_{NP}$ of the order of the weak scale or larger, only the
ultraviolet behaviour of the theory is modified, while the infrared
one is unaffected. This means that information about new
physics can be completely included in the matching conditions of the
Wilson coefficients, while the QED renormalization described by the
anomalous dimension matrix through eq.~(\ref{rge}) remains the same.

Furthermore, it is reasonable to assume that new physics has a chance to
affect sizably only matching conditions of operators that, in the SM,
are not generated at the tree level, but possibly at the quantum level.
This is indeed what happens in most of the SM extensions generally
considered. In this case, the analysis is particularly simple and
predictive.

Let us first consider new physics contributions to $\amu$. Following our
assumption, we expect that only the matching condition of the operator
$H_\mu$ is affected, but not those of the four-fermion operators
$V$ and $A$. This means that the total $\amu$ is just given by the sum of the
SM result discussed in the previous section and of a new contribution
$\amu^{\rm NP}$. If $\amu^{\rm NP}$ is known at one loop, the leading
contribution at two loops, enhanced by large QED logarithms, can be simply
obtained
\beq
a_\mu^{\rm NP} ({\rm 2\,loop})= 
-\frac{4\alpha}{\pi} \ln \frac{\Lambda_{NP}}{m_\mu} ~
a_\mu^{\rm NP} ({\rm 1\,loop})~.
\label{bubba}
\eeq
The coefficient in eq.~(\ref{bubba}) corresponds to the anomalous dimension
element $\gamma (H_\mu , H_\mu)$.
For instance, for $\Lambda_{NP}=100$ GeV, the inclusion of the
two-loop contribution reduces the one-loop result by 6{\%}.
This result is quite model independent, and it can be applied
to specific models in which $a_\mu^{\rm NP} ({\rm 1\,loop})$ is known, like
in the case of supersymmetry~\cite{susy}, light-gravitino 
interactions~\cite{gra}, compositeness~\cite{comp}, leptoquarks~\cite{lept},
and light non-minimal Higgs bosons~\cite{hig}.

Another case in which the QED logarithms turn out to be large is represented
by the EDM of the electron, $d_e$. In the SM $d_e$ is negligible, but in 
some of its extensions with new sources of CP violation, it can lie just
below the present experimental upper bound. For instance, this can be
the case in some versions of the supersymmetric model~\cite{sedm}.

The analysis of the QED renormalization can be done along the same lines
followed for $\amu$. The electron EDM corresponds to an operator in the 
effective Lagrangian
\beq
-\frac{i}{2}d_e ~\bar e \,\sigma_{\nu \rho}\gamma_5 \, e~ F^{\nu \rho}.
\label{edmop}
\eeq
It is easy to show that in the effective theory below the weak scale, its 
one-loop anomalous dimension is equal to that of the magnetic dipole 
operator that is given in eq.~(\ref{ano1}).
Therefore if $d_e^{(0)}$ describes the new physics contribution
obtained by integrating out the heavy modes with mass $\Lambda_{NP}$, the
QED improved result is
\beq
d_e = d_e^{(0)}\left[ 1-\frac{4\alpha}{\pi} \ln \frac{\Lambda_{NP}}{m_e}
\right] ~.
\label{edm}
\eeq
For $\Lambda_{NP}=100$ GeV, the term inside brackets amounts to a
reduction of $d_e^{(0)}$ of 11{\%}.

The result in eq.~(\ref{edm}) is rather model independent. However, it
should be remarked that it is valid only if $\Lambda_{NP}$ is not much
larger than the weak scale. An analysis at very large scales should 
include the mixing of the operator in eq.~(\ref{edmop}) with other
CP violating operators, like the analogue of eq.~(\ref{edmop}) for
the different electroweak gauge bosons, and the analogue of the
Weinberg operator~\cite{wei} for the $SU(2)$ gauge theory. The 
corresponding anomalous
dimension matrix can be extracted from 
ref.~\cite{brat}. The result however is more involved than the
one presented in eq.~(\ref{edm}), since it depends on the separate 
unknown coefficients of the different operators.

We wish to thank G. Altarelli and M. Ciuchini for useful discussions.

\def\ijmp#1#2#3{{\it Int. Jour. Mod. Phys. }{\bf #1~}(19#2)~#3}
\def\pl#1#2#3{{\it Phys. Lett. }{\bf B#1~}(19#2)~#3}
\def\zp#1#2#3{{\it Z. Phys. }{\bf C#1~}(19#2)~#3}
\def\prl#1#2#3{{\it Phys. Rev. Lett. }{\bf #1~}(19#2)~#3}
\def\rmp#1#2#3{{\it Rev. Mod. Phys. }{\bf #1~}(19#2)~#3}
\def\prep#1#2#3{{\it Phys. Rep. }{\bf #1~}(19#2)~#3}
\def\pr#1#2#3{{\it Phys. Rev. }{\bf D#1~}(19#2)~#3}
\def\np#1#2#3{{\it Nucl. Phys. }{\bf B#1~}(19#2)~#3}
\def\mpl#1#2#3{{\it Mod. Phys. Lett. }{\bf #1~}(19#2)~#3}
\def\arnps#1#2#3{{\it Annu. Rev. Nucl. Part. Sci. }{\bf
#1~}(19#2)~#3}
\def\sjnp#1#2#3{{\it Sov. J. Nucl. Phys. }{\bf #1~}(19#2)~#3}
\def\jetp#1#2#3{{\it JETP Lett. }{\bf #1~}(19#2)~#3}
\def\app#1#2#3{{\it Acta Phys. Polon. }{\bf #1~}(19#2)~#3}
\def\rnc#1#2#3{{\it Riv. Nuovo Cim. }{\bf #1~}(19#2)~#3}
\def\ap#1#2#3{{\it Ann. Phys. }{\bf #1~}(19#2)~#3}
\def\ptp#1#2#3{{\it Prog. Theor. Phys. }{\bf #1~}(19#2)~#3}

\end{document}